\documentclass[letterpaper]{article}

\usepackage{natbib,alifeconf} 
\usepackage{amssymb}
\usepackage{url}

\title{Innovation and informal knowledge exchanges between firms}

\author{Juste Raimbault$^{1,2,3,4}$\\
\mbox{}\\
$^1$LASTIG, Univ Gustave Eiffel, IGN-ENSG, Saint-Mand{\'e}, France\\
$^2$Centre for Advanced Spatial Analysis, UCL, London, United Kingdom\\
$^3$UPS CNRS 3611 ISC-PIF, Paris, France\\
$^4$UMR CNRS 8504 G{\'e}ographie-cit{\'e}s, Paris, France\medskip\\
juste.raimbault@polytechnique.edu}

\begin{document}
\maketitle

\begin{abstract}
Firm clusters are seen as having a positive effect on innovations, what can be interpreted as economies of scale or knowledge spillovers. The processes underlying the success of these clusters remain difficult to isolate. We propose in this paper a stylised agent-based model to test the role of geographical proximity and informal knowledge exchanges between firms on the emergence of innovations. The model is run on synthetic firm clusters. Sensitivity analysis and systematic model exploration unveil a strong impact of interaction distance on innovations, with a qualitative shift when spatial interactions are more intense. Model bi-objective optimisation shows a compromise between innovation and product diversity, suggesting trade-offs for clusters in practice. This model provides thus a first basis to systematically explore the interplay between firm cluster geography and innovation, from an evolutionary perspective.
\end{abstract}


\section{Introduction}

Innovation is a central process of evolution, from biological evolution to social, cultural \citep{mesoudi2018cumulative} and technological evolution \citep{sood2005technological}. Understanding the drivers of technological innovation is in that context crucial from a theoretical perspective for insights into evolutionary theories of social change and evolutionary economics among others, and from a practical perspective for sustainable planning and management of societies. Technological innovation may indeed be an essential aspect of transitions towards sustainability \citep{adams2016sustainability}, although it should not be their sole driver at the detriment of other dimensions of transitions such as social change.

Geographical proximity, or in practice the implementation of firm clusters, is thought to have a positive impact on innovation capabilities \citep{bittencourt2019cluster}. In that context, the role of local informal interactions between innovation agents has been suggested as important for breakthrough innovations by empirical and theoretical studies. In the context of firm cluster, \cite{gnyawali2013complementary} propose the intensity of social interaction as a key factor alongside cluster competition intensity to determine potential future innovations. Clusters are understood as enablers of tacit knowledge exchanges between inventors from different firms \citep{arikan2009interfirm}. Furthermore, the mobility of employees between firms in the same area may be a support for the transfer of competences and tacit knowledge \citep{almeida1999localization}. Firms benefit from a stronger connection in local social networks \citep{kemeny2016economic}. The idea of firms as innovation incubators in which ideas evolve can be linked to an evolutionary approach to social systems which has been widely studied by the Artificial Life community \citep{marriott2018social}.

From an evolutionary perspective, the concept of market niche has been used to explain technological change \citep{schot2007niches}. The firm in that context acts as the primary space where evolution of ideas occurs, and knowledge flows between firms can be understood in analogy with gene flows between isolated geographical areas in biological innovation. The transfer of concepts from biology to economic geography remains however valid only to a certain extent \citep{schamp2010notion}, and a precise definition of genomes, species, evolution and co-evolution in social systems is not straightforward \citep{raimbault2019modeling}. Regarding innovation, multiple scales from firms to cities can be for example considered \citep{raimbault2020model}. We choose here to focus on the microscopic scale of innovation ecosystems, more precisely how research and development employees of firms act as carriers of ideas leading to the emergence of breakthrough innovations \citep{song2016innovation}.

One privileged tool to study and simulate the emergence of innovations from this microscopic perspective are agent-based models (ABM). Various ABMs have been proposed for the diffusion of innovation \citep{kiesling2012agent}. \cite{sayama2015studying} use ABMs to simulate an ecology of ideas and study collective decision making and creativity. \cite{lopolito2013emerging} combine knowledge exchange, expectations of agents and learning as core mechanisms to simulate innovation niches. \cite{dosi2021patents} introduce an ABM to investigate the role of patenting on the innovativity of firms competing on a set of submarkets, including consumer demand. \cite{chen2006functional} focus on functional modularity of products and use a genetic programming formalism to evolve technologies. \cite{ma2005agent} describe an ABM of technological change which takes into account both intrinsic fitness selection pressure and environment selection pressure, the latest being determined by the interaction with customers. The role of space is studied by \cite{vermeulen2018role} in a multi-level approach combining interregional knowledge networks and knowledge diversity within regions. Diverse aspects of firm clusters have been studied by means of ABMs, such as firm competitiveness, local networks, and policy-making, among others \citep{fioretti2005agent}.

These previous modeling efforts however do not specifically tackle the particular question of informal knowledge flows within firm clusters. It remains still an important dimension, with implications in urban planning and concrete aspects of firm cluster implementation among others. We propose in this paper to study this issue by developing a stylised agent-based model of technological change within firm clusters. Following the approach of artificial societies \citep{epstein1997artificial}, we do not aim at providing a highly realistic or data-driven model, but rather a simple tool to explore the interplay between basic mechanisms in the emergence of a macroscopic phenomenon (innovation within firms in our case). More precisely, our contribution relies on the following points: (i) we provide a simple ABM linking innovation within firms and informal knowledge flows within firm clusters, based on an evolutionary model for innovation and exhibiting a strong analogy with biogeography optimisation algorithms \citep{simon2008biogeography}; (ii) the model is implemented with a specific instance for the genotype-phenotype mapping, using a generalised Rastrigin function as synthetic fitness landscape; (iii) the model is systematically explored, using global sensitivity analysis and a genetic algorithm optimisation to unveil various patterns of innovation in firm clusters.

The rest of this paper is organised as follows: we first describe and formalise the agent-based model; we then describe results of various numerical experiments; and finally discuss the implications of these results and diverse potential model developments and applications.

\section{Agent-based model}

\subsection{Rationale}

The main structure of the model corresponds to a set of firms, each composed by a set of employees. An employee is represented by some ideas, and these are mixed through evolution crossover within firms, but also mutated at the scale of each individuals. Indeed, empirical evidence using patents as a proxy of innovation suggest that inventions are produced by the superposition of exploration (recombination of existing technologies, which would correspond to a crossover in the genome) and exploitation (small incremental changes to existing combinations, captured by a local gene mutation) processes \citep{youn2015invention}. 

The main feature of the model is an additional crossover between firms, which captures the process of informal knowledge flows. In practice, employees of different firms in the same sector living in the same geographical area will share connections through social networks, meet intentionally or unintentionally, and share ideas. Although in many cases professional secrecy is strictly observed, a tension with knowledge sharing exists \citep{rouyre2019managing}. Therefore, informal knowledge is still exchanged, on non-sensitive subjects such as work or management practices, or technical subjects unrelated to the company's core business. We model interactions between employees of different firms through spatial interaction modeling \citep{wilson1975some}, which has already been used to study innovation and knowledge spillovers \citep{lesage2007knowledge}. In practice, adding this geographical component makes our model closer to biogeography optimisation algorithms \citep{simon2008biogeography}.

At the scale of intra-firm innovation, we need to introduce an evolution model. Therefore, it is necessary to specify a selection process linked to some mapping between the genotype of inventions and their phenotype, in other words a fitness function. \cite{ma2005agent} use a linear mapping obtained by applying a constant matrix to the genome, inspired from the model of \cite{kauffman1995technological}. The concept of fitness landscape is applied in different streams of complexity economics \citep{khraisha2020complex}. We choose to work with a similar heuristic, using a complicated fitness landscape obtained from the genome. Our model is in practice applicable with any optimisation landscape, but for the sake of simplicity we will work below with a generalised Rastrigin function which is a classical difficult optimisation problem used as a benchmark for optimisation algorithms.

\subsection{Model description}

The core element of the agent-based model are firms $f_k$ with $1 \leq k \leq N_f$. These are located in space by coordinates $(x_k,y_k) \in \mathbb{R}^2$. Each are composed by a set of employees $e_{ki}$ with $1 \leq i \leq S_k$ where $S_k$ is the size of the firm. In principle, number of firms, locations and sizes can take any value, but we will parametrise them with realistic values as detailed later. An employee is summarised by a set of ideas, captured by a real genome of fixed size: $e_{ki} = (x^{(ki)}_j) (t) \in \mathbb{R}^G$ where $G$ is the genome size. These ideas will evolve as time step $t$ changes. We do not include more detailed employee characteristics such as home location, assuming that spatial interaction modeling captures microscopic interactions around firms. We also do not include competences or field, as the innovation model is a simple genetic algorithm without detailed economic structure. At a given time step, a firm is also characterised by its current product $p_k (t) = (p_{kj})(t)\in \mathbb{R}^G$ and the corresponding fitness value $y_k (t)$ (which can be interpreted as a societal value of the innovation, or as the turnover potential of the product for the firm).

Starting from an initial state at $t = 0$, the model proceeds iteratively to evolve and exchange ideas, and to innovate within firms, for a fixed number of time steps until final time $t_f$. At each time step, the following actions are taken in order.

\begin{enumerate}
    \item Ideas are exchanged within firms, corresponding to the core of the genetic algorithm capturing innovation. Each employee has a fixed probability $p_C$ of realising a crossover with another one. Following a random draw, if this is realised, one other employee is selected at random, and the current employee copies a fixed share $s_C$ of the other genome (obtained in practice with a random draw of probability $s_C$ for each gene). The update is done synchronously so that no propagation can occur within one time step. The exchange is not symmetric (reflecting the asymmetry of idea exchanges in real life), but each employee gets to renew its ideas. Genomes are then mutated with a probability $p_M$ (at the gene level for all employees), and mutations correspond to a uniform random increment $m \in \left[ -x_M/2 ; x_M/2 \right]$ where $x_M$ is a parameter giving the amplitude of the mutation.
    \item New ideas are tried by employees, in other words the fitness function $y$ is evaluated for each genome $y_{ki} = y(e_{ki})$. Within each company, the next product is selected as the one maximising the fitness: $p_k (t) = \textrm{argmax}_i y_{ki}$ and the corresponding fitness value is taken as $y_k (t)$. At this stage, the analogy with the genetic algorithm is slightly modified to reflect actual research processes within firms: a fixed share of employees $s_P$ of the firm is randomly chosen to work on the product during the next cycle, and thus update their genome as the product genome $e_{ik} = p_k$. This leads to a loss of diversity which may be detrimental to a genetic algorithm with the sole aim of optimisation, but this is not the purpose of our model which is to capture actual innovation processes in a stylised way.
    \item Informal knowledge flows occur between firms, in practice being carried by local social networks of employees and their daily activities within the geographical area of the cluster. We assume informal flows within firms are indistinguishable from formal work exchanges accounted for at the first step, and this step only captures inter-firms interactions. For any pair of employees $(e_{k_i i}, e_{k_j j})$ from distinct firms $k_i,k_j$, a probability of interacting is given by
    \[
    p_{ij} = p_E \cdot \exp{\left( - d(k_i,k_j) / d_E \right)}
    \]
    where $p_E$ is a parameter giving the local intensity of informal exchanges (which will for example capture the difference between a rural, periurban and urban cluster), $d(k_i,k_j)$ is the geographical distance between firms and $d_E$ is the characteristic distance of interactions. Taking the average on employees and aggregating by firms gives the expected number of interactions between firms as
    \[
    I_{kl} = S_k \cdot S_l \cdot \exp{\left( - d(k,l) / d_E \right)}
    \]
    which corresponds to a classic spatial interaction model \citep{wilson1975some}. Two interacting employees will act as for the internal crossover: the first agent copies a random part of the genome using the $s_C$ parameter.
\end{enumerate}

\subsection{Model indicators}

We study various aspect to quantify model dynamics and outcomes. First, innovation utility is measured through fitness values. At each time step, we consider (i) $b(t) = \max_k y_k (t)$ the best fitness value across all firms, and (ii) $\bar{f} (t)$ the average fitness value across firms. We then capture economic inequality between firms, both through the relative fitness difference $\Delta f$ between the best and the worst performing company, and with the entropy $\mathcal{E}_f$ of the distribution of fitness. Finally, we capture product diversity at the genotype level (which is complementary to previous inequality indicators at the phenotype level), using an average dissimilarity index obtained with cosine similarity:

\[
d(t) = \frac{1}{2 \cdot N_f \cdot (N_f - 1)} \sum_{k \neq l} \left(1 - \frac{p_k (t)\cdot p_l (t)}{\left||p_k (t)\right|| \cdot \left||p_l (t)\right||}\right)
\]

The more diverse phenotypes are, the higher this diversity index will be.

\subsection{Model setup}

A certain number of parameters can be parametrised to match realistic configurations. The size of firms can in practice be approximated by a power law for the largest sizes of the distribution \citep{growiec2008size}. Therefore, we distribute $S_k$ following a rank-size law $S_k = S_0 \cdot k^{-\alpha_S}$ where indices are in decreasing size order, $S_0$ is the size of the largest firm and $\alpha_S$ the level of hierarchy. Locations of firms are taken randomly in $\left[0 ; 100 \right]^2$, such that the order of magnitude for the $d_E$ is in the same interval (corresponding to realistic sizes of urban regions where clusters are generally located). Initial employee genomes are initialised randomly in $[-10 ; 10]$ and the initial product and corresponding fitness are chosen randomly among employees.

Regarding the fitness landscape, we work on an particular implementation using a function difficult to optimise. We work with a generalised Rastrigin function, that we define here as
\[
y(\vec{x}) = - \sum_{i,j} m_{ij} \left[x_i^2 - 10 \cos\left(2 \pi x_i\right) \right]
\]
where $m_{ij}$ is a random uniform static matrix of size $G \times G$ and with coefficient in $[0;1]$, capturing the random fitness landscape used by  \cite{ma2005agent}, and the rest is the classic Rastrigin function.

We furthermore fix a certain number of parameters which can reasonably correspond to real world values. We take medium-sized companies by taking $S_0 = 100$ and a number of firms $N_f = 10$, corresponding to a medium-sized cluster, such as in the case of several start-ups working in digital services. We fix the genome size at $G = 10$ to avoid exploring too large dimensional spaces. We run the model with $t_f = 100$, corresponding to a magnitude of 10 years if one time steps is roughly one month. The rest of the parameters are left free and will be explored in the numerical experiments.

\section{Results}

\begin{table*}[ht!]
\caption{Saltelli sensitivity indices, for indicators at $t_f$ in rows and parameters in columns. We give for each pair the first order index (F) and the total order index (T). Non-significant values were assimilated to 0.\label{tab:gsa}}
\resizebox{\linewidth}{!}{  
\begin{tabular}{|l|c|c|c|c|c|c|c|c|c|c|c|c|c|c|c|c|c|c|}
\hline
 & \multicolumn{2}{|c|}{$\alpha_S$} & \multicolumn{2}{|c|}{$p_C$} & \multicolumn{2}{|c|}{$s_C$} & \multicolumn{2}{|c|}{$p_M$} & \multicolumn{2}{|c|}{$x_M$} & \multicolumn{2}{|c|}{$s_P$} & \multicolumn{2}{|c|}{$p_E$} & \multicolumn{2}{|c|}{$d_E$} & \multicolumn{2}{|c|}{seed}  \\
 & F & T & F & T & F & T & F & T & F & T & F & T & F & T & F & T & F & T \\
 \hline
$b$ & 0.001 & 0.002 & 0.001 & 0.003 & $9\cdot 10^{-4}$ & 0.002 & 0.41 & 0.75 & 0.17 & 0.52 & 0.03 & 0.13 & $5\cdot 10^{-4}$ & 0.002 & $9\cdot 10^{-4}$ & 0.002 & 0.003 & 0.007\\
$\bar{f}$ & 0.02 & 0.07 & $6\cdot 10^{-4}$ & 0.002 & 0.0 & 0.003 & 0.36 & 0.69 & 0.21 & 0.55 & 0.02 & 0.008 & 0.0 & 0.004 & $4\cdot 10^{-4}$ & 0.004 & $8\cdot 10^{-4}$ & 0.007\\
$\Delta f$ & $7\cdot 10^{-4}$ & 0.56 & 0.0 & 0.9 & 0.0 & 0.0 & 0.003 & 0.0 & 0.0 & 0.24 & 0.0 & 0.48 & 0.0 & 0.18 & 0.0 & 0.17 & 0.0 & 0.0 \\
$\mathcal{E}_f$ & 0.14 & 0.64 & 0.0 & 0.44 & 0.27 & 0.36 & 0.48  & 0.84 & 0.014 & 0.35 & 0.23 & 0.41 & 0.16 & 0.39 & 0.0 & 0.40 & 0.05 & 0.46 \\
$d$ & 0.007 & 0.13 & 0.001 & 0.04 & 0.01 &  0.1 &  0.45  & 0.7 & 0.21 & 0.42  & 0.0  & 0.1 & 0.003 & 0.09 & 0.006 &  0.09 & 0.006 & 0.05 \\\hline
\end{tabular}
}
\end{table*}

The model is implemented in \texttt{scala} for performance purposes. Simulations and design of experiments are achieved using the software OpenMOLE for model exploration and validation \citep{reuillon2013openmole}, which provides seamless model embedding, simple access to high performance computing infrastructures and state-of-the-art model sensitivity analysis and validation techniques. Model source code is open and available on the git repository of the project at \url{https://github.com/JusteRaimbault/InnovationInformal}. Simulation results used in the paper are available on the dataverse repository at \url{https://doi.org/10.7910/DVN/X8PWPF}.

Explored parameter space corresponds to, when not specified otherwise, the following parameters and ranges: firm size hierarchy $\alpha_S \in [0.1 ; 2.0]$, crossover probability $p_C \in [0 ; 1]$, crossover share $s_C \in [0 ; 1]$, mutation probability $p_M \in [0 ; 1]$, mutation amplitude $x_M \in [0 ; 2]$, product work share $s_P \in [0 ; 1]$, interaction probability $p_E \in [0 ; 10^{-4}]$ (this highest bound gives already a considerable mixing of ideas leading to a total uniformity of products), and distance decay $d_E \in [1 ; 100]$.

\begin{figure*}
\begin{center}
\includegraphics[width=0.9\linewidth]{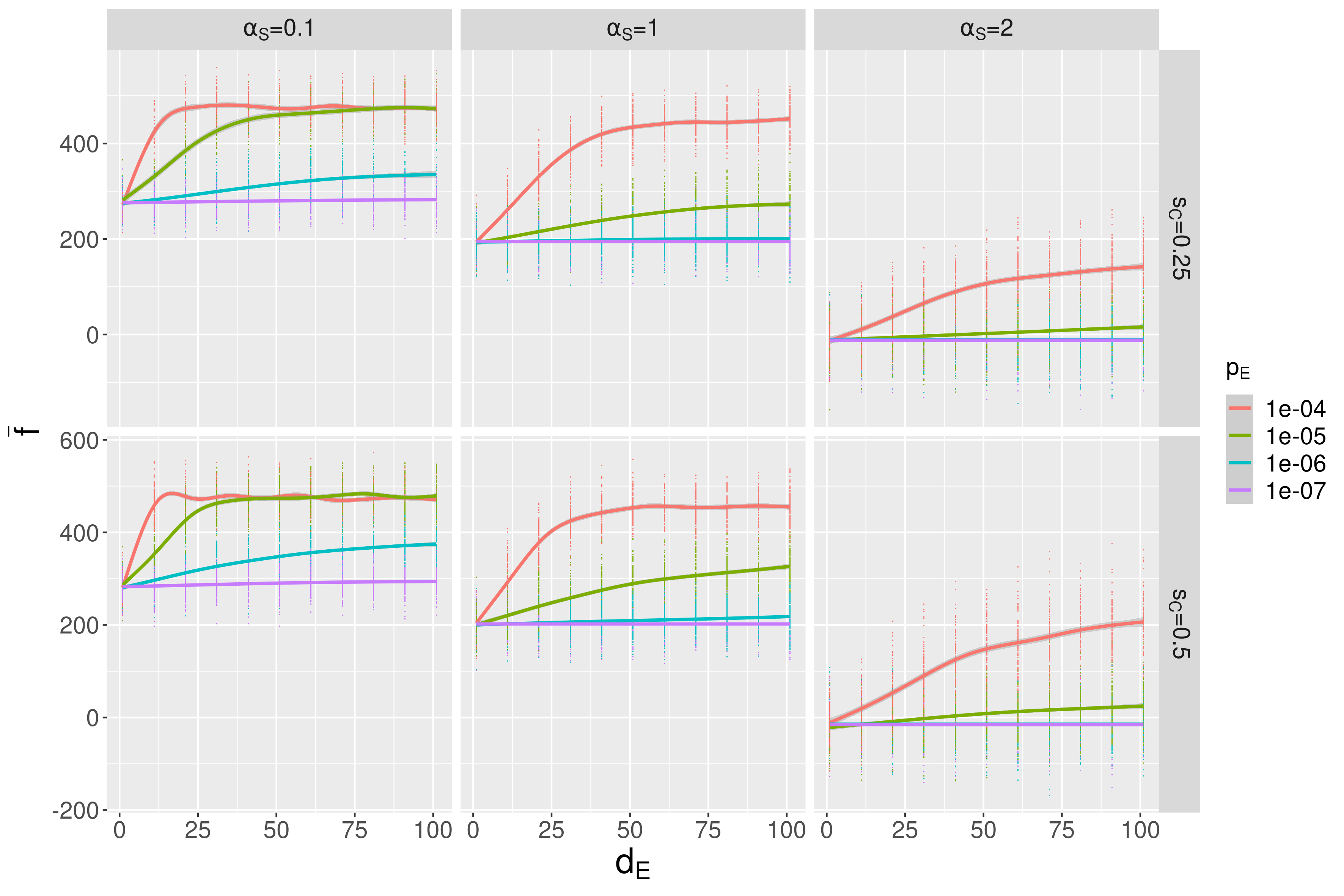}\\
\includegraphics[width=0.9\linewidth]{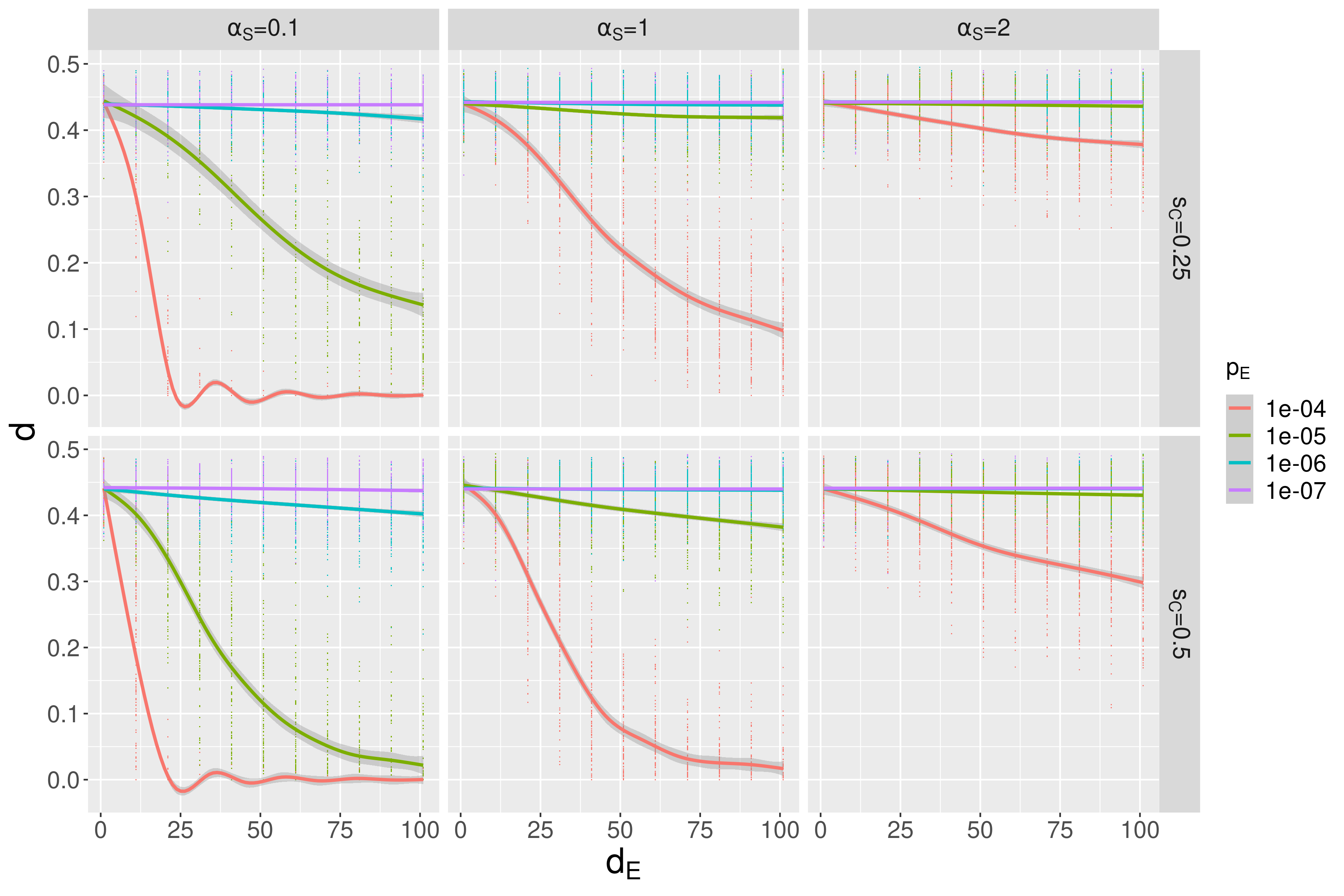}\\
\caption{Behavior of average fitness (top plot) and product diversity (bottom plot), as a function of distance decay $d_E$. We plot raw replication points and smoothed average values, for different values of interaction probability (color scale), and for varying firm size hierarchy $\alpha_S$ (columns) and crossover share $s_C$ (rows). Both plots are shown for $p_C = 0.5$, with no significant qualitative change for $p_C = 0.25$.}
\label{fig:fig1}
\end{center}
\end{figure*}

\begin{figure*}[h!]
\begin{center}
\includegraphics[width=0.48\linewidth]{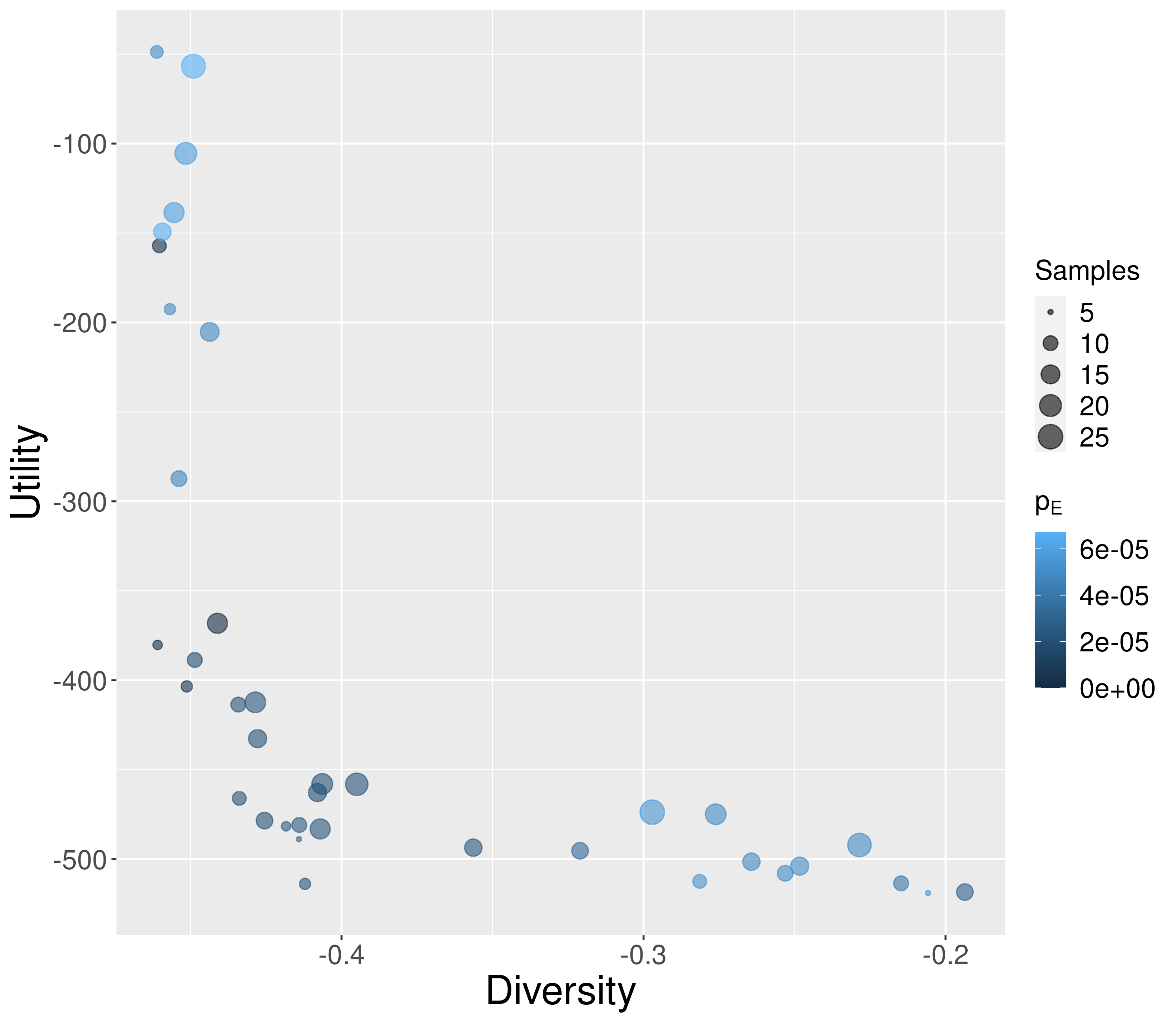}
\includegraphics[width=0.48\linewidth]{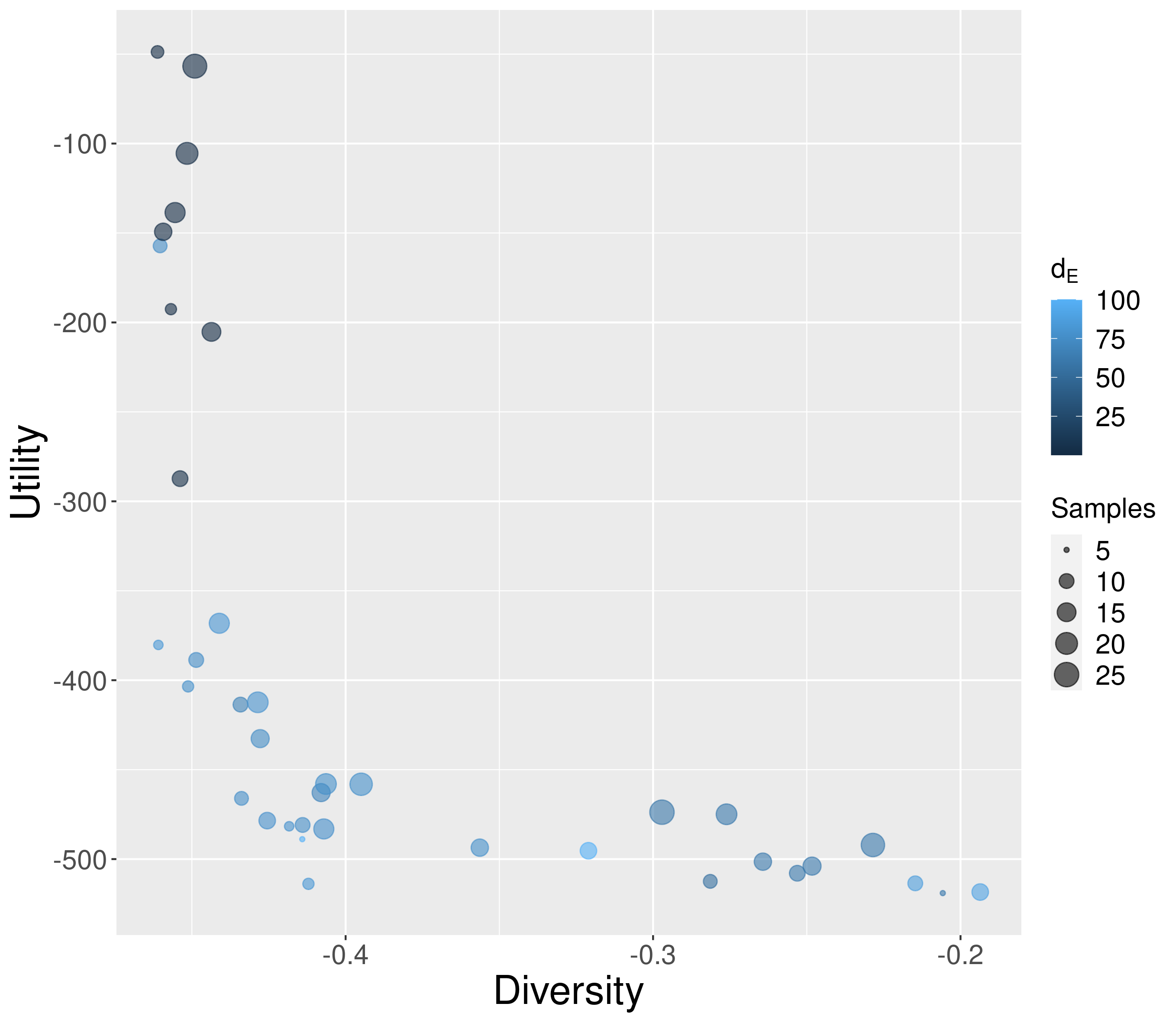}
\caption{Pareto fronts between average fitness and diversity, obtained with a NSGA2 bi-objective optimisation algorithm. Point size gives the number of stochastic samples, while point color gives interaction probability $p_E$ for the left plot and distance decay $d_E$ for the right plot.}
\label{fig:fig2}
\end{center}
\end{figure*}

\subsection{Statistical convergence}

We first proceed to an internal validation experiment, aimed at testing whether model outputs are robust to stochasticity, or in other words if they exhibit good statistical convergence properties. We sample 100 parameter points using a Latin Hypercube Sampling \citep{giunta2003overview}, and run 1000 replications of the model for each parameter point. This large number of replications is first necessary to estimate the statistical properties of indicators.

We first look at the variability of indicators, looking at Sharpe ratios defined as $\mu \left[ I \right] / \sigma \left[I \right]$ which $\mu$ and $\sigma$ estimators of mean (resp. variance) for the indicator $I$. Most indicators, including best and average fitness, fitness entropy and diversity, exhibit a low variability with the first quartile of Sharpe ratios larger than 4 across all parameter points. Fitness relative difference is more stochastic with a median of 1.48. Altogether, indicators have thus a low variability.

We then investigate how to discriminate two estimated average indicator values. A relative distance between averages is given by $2 \cdot \frac{\mu \left[ I \right] + \mu \left[ J \right]}{\sigma \left[I \right]+ \sigma \left[J \right]}$, for indicators $I,J$ and estimated across all pairs of parameter points. This value is high for best and average fitness (first quartile at 6.8 and 9.9), low for inequalities (median at 2.5 for relative difference and 0.56 for entropy), and relatively high for diversity (first quartile at 2.8). In the case of normal distributions, a confidence interval of size $\sigma/2$ is obtained with 64 repetitions (as confidence interval size is $2\cdot 1.96\cdot \sigma / \sqrt{n}$), so we run our experiments with $n=100$ to ensure a proper separation of indicator values.

\subsection{Global sensitivity analysis}

We then proceed to a global sensitivity analysis to investigate the respective role of parameters in terms of indicators variance. This technique described by \cite{saltelli2008global} provides aggregate measures of parameter relative importance, both at the first order (all other parameters being fixed) and also capturing interaction effects with other parameters (total order). We take $N = 10,000$ design points for the estimation. Results of sensitivity indices estimation are shown in Table~\ref{tab:gsa}. We find that the size distribution of firms influences fitness inequality and diversity, but not performance. Crossover parameters mostly influence the entropy of fitness. The parameter with most influence overall is mutation probability, with 3 indicators being significantly changed. We will fix this parameter in the following to ensure a refined exploration, focusing on second order effects. Share of product adoption within the firm $s_P$ has a significant total order influence on best fitness, what may correspond to the fact that this parameter sometimes induces innovation locks through the loss of diversity. Spatial interaction parameters $p_E$ and $d_E$ influence strongly inequality but not diversity, although some diversity reduction could have been expected from exchanges. Finally, we confirm the low sensitivity to stochastic noise as all indicators have low indices with respect to the random seed. Altogether, this global sensitivity analysis confirms that all mechanisms play a role and that they interact in a complex manner.

\subsection{Parameter space exploration}

We now turn to a more targeted exploration of the parameter space to discuss model behavior. We choose to fix mutation parameters in order to focus on the role of exchanges and geography. We therefore take $p_M = 0.01$, $x_M = 1$. We also fix current product share, as it is similarly a specific parameter of the genetic algorithm, and in terms of thematic interpretation is internal to firms. We take an intermediate value of $s_P = 0.5$. The crossover parameter $s_C$ is in contrary involved in informal knowledge exchanges. We explore a coarse grid for $s_C \in \{0.25 ; 0.5\}$, for $p_C \in \{0.25 ; 0.5\}$ and for $\alpha_S \in \{0.1 : 1.0 : 2.0\}$, combined to a more refined grid for exchange parameters: $\log (p_E) \in \{-7 ; -6 ; -5 ; -4\}$ and $d_E$ varying from 1 to 101 with a step of $10$. We run 100 model replications for each parameter point, corresponding to a total of 52,800 model runs.

We show the behavior of main indicators in Fig.~\ref{fig:fig1}. Both average fitness and product diversity exhibit a similar behavior. When distance decays increases, i.e. when the integration of the firm cluster in terms of informal knowledge is strengthened, innovation is improved as the average fitness increases significantly. This effect disappears for low interaction probabilities, recalling that the informal knowledge flow is the outcome of these two processes of social intensity and spatial interactions. The increase in fitness is at the detriment of product diversity. When comparing columns with varying $\alpha_S$, we find that unequal firm sizes (larger $\alpha_S$ values) are non optimal, as highest fitness values are obtained for the lowest hierarchy. The crossover share $s_C$ (plot rows) does not change much indicator behavior, except for diversity at high interaction probabilities (middle column of bottom plot).

It is interesting to note that changing $p_E$ leads to a qualitative change in model regime: low values mostly imply a steady increase of fitness (decrease of diversity), while intense interaction lead to a sharp increase followed by a plateau. This implies a change in the way to conceive clusters depending on their geographical situation: proximity will be beneficial more quickly in urban environments compared to rural settings for example.

\subsection{Optimisation}

We finally run a bi-objective optimisation algorithm, to investigate the potential compromise between innovation in terms of fitness and product diversity. Indeed, diversity is crucial to maintain for a longer term robustness and resilience of the socio-technical system \citep{reinmoeller2005link}. We use a NSGA2 genetic algorithm with two optimisation objectives to be maximised: diversity and average fitness. We use the OpenMOLE implementation of a steady state NSGA2, with a population of 200 individuals, for 10,000 generations. In practice, this implementation minimises objectives, so we use opposites $- \bar{f}$ and $-d$ as optimisation objectives. The number of stochastic samples for each parameter point is determined through an embedding strategy, adding this number as an additional optimisation objective. In the final population, we filter points with less than 5 repetitions.

We show the optimisation results in Fig.~\ref{fig:fig2}. We find a Pareto front between the two objectives, confirming the compromise between global performance and diversity. Both extremities of the front are rather steep/flat, meaning that a reduced number of points provide an effective compromise. Investigating the values of some parameters, we find some kind of U-shaped behavior for interaction probability $p_E$: high values for this parameter put the points on extremities of the front. This is an interesting behavior as quite unexpected from an intuitive point of view: increasing interactions should mix more ideas and decrease diversity - which is true, but also includes the opposite, i.e. optimal diversity when interaction are high. The explanation may rely on the fact that these points (top-left extremity of the front) correspond to low values of distance decay $d_E$, as seen on the right plot of Fig.~\ref{fig:fig2}. The localised regime impedes the effect of interactions in that case. We observe a similar behavior with product share $s_P$, but with different underlying processes: a too high share could have been expected to induce technological locks. This shows altogether the complexity of interacting processes within firm clusters, leading to the emergence of innovations.

We can investigate the part of the Pareto front which would constitute some ``reasonable'' compromise, i.e. where trade-offs between the two objectives are of similar amplitude. We therefore filter the points such that $- \bar{f} < -400$ and $-d < -0.4$, obtaining 13 compromise points. Interestingly, the parameter values for these points are rather localised. Their values with standard deviations are: $\alpha_S = 0.13 \pm 0.04$, $p_C = 0.94 \pm 0.05$, $s_C = 0.23 \pm 0.05$, $p_M = 0.03 \pm 0.008$, $x_M = 1.26 \pm 0.34 $, $s_P = 0.12 \pm 0.05$, $p_E = 2 \cdot 10^{-5} \pm 5 \cdot 10^{-6}$ and $d_E = 77 \pm 7.8$. This corresponds to equal firm sizes, frequent crossovers of a quarter of the genome, very low mutations (as fixed in the grid sampling experiment), a small but not negligible product share (keeping diversity within companies is thus important for the compromise), and a very low interaction probability but at a long range. In practice, this would be interpreted a regional firm system with few but important informal idea exchanges between firms.

\section{Discussion}

We explored a novel model for innovation diffusion within and between firms from an evolutionary perspective. One important contribution of this work compared to previous literature is the stylised realistic parametrisation, coupling paradigms from evolutionary computation and economic geography. The main takeovers drawn from our simulations are (i) a strong effect of informal knowledge exchanges on innovation fitness, but which is rapidly plateauing; (ii) a more optimal configurations in terms of fitness when firms are close in size, compared to highly hierarchical firm systems; (iii) a compromise between innovation fitness and diversity, with the trade-off region of similar amplitude being characterised by an equal-size regional firm system. The second point relates with the idea of modular systems being a favourable context for innovation and creativity as found by \cite{dionne2019diversity}. In our model, spatial clusters corresponding to firms play a crucial role. The third point suggests that these clusters are balanced and geographically distributed in the compromise configuration. More generally, the configuration of spatial niches may play an important role in evolutionary systems.

Our stylised results can furthermore be linked with documented empirical facts. The innovation success of a firm cluster relies on a strong interplay between local interactions and global integration \citep{fitjar2014local}. Put in another way, local interactions are not sufficient to drive innovation. However, given the sharp fitness increase exhibited by our model when increasing local knowledge flows, we can suggest that these may be necessary, and that a firm in complete isolation would have difficulties to innovate (whether knowledge flows can occur without being local or informal is another question out of the scope of our work). Furthermore, although empirical stylised facts are not unanimously agreed on, there exists evidence that cluster size may lead to ``agglomeration diseconomies'', in other words that clustering becomes detrimental above a certain size \citep{folta2006geographic}. We do not obtain this aspect, since the effect of distance and interaction probability are always increasing in terms of fitness. However, the opposite effects on diversity may be interpreted as detrimental as maintaining diversity is important for the resilience of complex systems \citep{fraccascia2018resilience}. Finally, in relation with cluster size, we find that firm size hierarchy $\alpha_S$ is to be minimal (firm of equal sizes) to obtain a higher fitness. This implies that clusters should not be dominated by large companies for a better innovation performance.

Numerous extensions and applications are open at this point. More advanced model validation procedures would bring further knowledge on its complex behavior: behavior search algorithms provide a feasible output space \citep{cherel2015beyond}, while the Calibration Profile algorithm can be applied to conditional optimisation along discrete axis for parameters of interest \citep{reuillon2015new}. The several stylised facts contained within the conceptual model introduced by \cite{gnyawali2013complementary} may be the basis for more general models which would need to reproduce these facts. The combination of this model with urban innovation diffusion models such as \citep{raimbault2020model} could provide a multiscale model of innovation clusters. Exploring other instances of fitness landscapes is also crucial to assess the robustness of our results and to be able to generalise. Regarding aspects that were not taken into account, teleworking can significantly change the role of informal knowledge exchanges and the geography of clusters. It was shown recently to influence the productivity of firms \citep{bergeaud4015066telework} and is one component of what \cite{duranton1999distance} calls the ``tyranny of proximity'': face-to-face contacts have a novel importance in that context. Furthermore, this model could be parametrised and calibrated on real world data, including patent data for innovation and real cluster case studies. Finally, this model could have potential policy applications, to plan and manage company clusters to foster innovation in the context of sustainability.

To conclude, we have introduced and explored a simple instance of an innovation diffusion model, focused on informal knowledge flows and the geography of firm clusters. The model was explored for a particular instance of fitness landscape. We find a strong effect of these flows on innovation performance, and a compromise between diversity and innovation corresponding to regional firm systems. These results and the model can be the basis of future empirical, theoretical and modeling research, in link with policy applications.

\end{document}